\documentclass[aps,twocolumn,prd,showpacs,floatfix]{revtex4}
\usepackage{bm}
\usepackage{amsmath}
\usepackage{amsthm}
\usepackage{amssymb}
\usepackage{graphicx}
\usepackage{nicefrac}
\usepackage{multirow}

\newcommand{\eref}[1]{(\ref{#1})}

\begin{document}

\title{Calculation of strongly forbidden M1 transitions and 
  g-factor anomalies in atoms considered for parity non-conservation measurements} 
\author{G. H. Gossel}
\affiliation{School of Physics, University of New South Wales, Sydney
  2052, Australia} 
\author{V. A. Dzuba}
\affiliation{School of Physics, University of New South Wales, Sydney
  2052, Australia} 
\author{V. V. Flambaum}
\affiliation{School of Physics, University of New South Wales, Sydney
  2052, Australia} 

\date{\today}

\begin{abstract}
We calculate magnetic dipole transition amplitudes in $s-s$ and $s-d$
transitions of Rb, Cs, Ba$^+$, Fr, Ra$^+$, Yb$^+$, Ac$^{2+}$ and
Th$^{3+}$. These transitions were used or considered to be used for the
measurements of parity non-conservation (PNC). We also calculate the magnetic g-factor anomalies for a selection of states, along with electric quadrupole transition amplitudes for $s-d$ transitions in these systems.

\end{abstract}


\maketitle  
\section{Introduction}
Currently the interest to study parity non-conservation (PNC) in atoms
remains high due to its status as the best low-energy test of the standard
model. There is interest in obtaining important information by improving
the accuracy of the measurements and their interpretation, studying
PNC in a chain of isotopes, and measuring nuclear P-odd anapole
moments (see, e.g. review ~\cite{rev}). The most accurate test of the standard
model in atomic PNC comes from the study of the $6s -7s$ PNC
transition amplitude in cesium~\cite{Wood}. This is due to the extremely
high accuracy of measurements~\cite{Wood} and
calculations ~\cite{DFG02,PBD09,DBFR}. The
$6s-7s$ transition in cesium is a strongly forbidden magnetic dipole
(M1) transition. Using the strongly forbidden  M1 transitions for PNC measurements
was first suggested by Bouchiat~\cite{Bouchiat}.
 More recent proposals to use M1
transitions for PNC measurements include $s-s$ and $s-d$ transitions
in Ba$^+$~\cite{Ba+}, Ra$^+$~\cite{Ra+}, Yb$^+$~\cite{Yb+},
Fr~\cite{FrPNC}, Rb~\cite{RbPNC}, and Fr-like ions~\cite{Fr-like} (see
also Ref.~\cite{s-d,BaYbRa,PorsevYb+}). 

The experimental data on the values
of the M1 transition amplitudes is poor. Among the mentioned atoms there
are only experimental data for the $6s - 7s$ transition in
cesium~\cite{CsM1a,CsM1b,CsM1c}, and plans to measure the $6s - 5d_{3/2}$ M1
amplitude in Ba$^+$~\cite{BaM1}. Knowing the value of this M1 amplitude is important when planning and interpreting the measurements and for testing atomic theory~\cite{BaM1}.  

In this work we present theoretical calculations of M1 transition
amplitudes in a variety of systems considered for PNC
measurements~\cite{RbPNC,s-d,Fr-like}. These
amplitudes are calculated in the relativistic Hartree-Fock
approximation with contributions from core polarization and the
Breit interaction included to all orders.
 We also calculate magnetic g-factor anomalies which are produced by the same mechanisms as the strongly forbidden M1 transitions and provide a good test of the accuracy. There are two main mechanisms  (see, e.g.~\citep{HeavyAtom1}). The first one is due to the relativistic corrections to the magnetic moment operator  ($\sim - \alpha^2=-(1/137)^2$) and dominates in light atoms. The second mechanism is due to the combined action of the exchange core polarization and the spin-orbit interaction. It increases very fast with the nuclear charge ($\sim  Z^4 \alpha^4$) and dominates in heavy atoms ~\citep{HeavyAtom1}. In the present work we use the relativistic magnetic moment operator, therefore
the relativistic corrections (the first mechanism) are included at the Hartree-Fock level. The second mechanism is included in the calculation of the core polarization effects.  
In many cases the core polarization is essentially the sole contributor
to the M1 amplitudes and g-factor anomalies (compared with Hartree-Fock and Breit alone). Comparison of the calculated M1 amplitude
may be made with experiment in the case of cesium, for which there is
good agreement.
 We also present results for other systems, including
$\mathrm{Ba^+}$ (for which PNC measurements are being
considered~\citep{BaII}) and $\mathrm{Ra^+}$ and Fr, for which PNC
measurements are underway~\citep{FrPNC,RaII}). Also presented are calculations of $s-d$ electric quadrupole (E2) transition amplitudes, including both core polarization and electron correlation effects. Calculated E2 values are compared with previous calculations for Ba$^+$, Ra$^+$, and Ac$^{2+}$. Experimental E2 data exists for Cs and Ba$^+$, and comparing with our calculated values shows good agreement for Ba$^+$, however the Cs data has poor experimental accuracy.

To estimate the accuracy of the M1 amplitudes calculations we present calculations of the  g-factor anomalies in the
same atoms and ions, as well as a comparison with the available experimental data. The deviation of the g-factor of $s$ states of
single-valence-electron atoms from the g-factor of a free electron,
known as the g-factor anomaly, has been considered in many atomic systems \citep{CsExp,CsExp2,CsRbGfactor,RbExp,RbExp2,FrExpG,RelCorrectionsG,HeavyAtom1,HeavyAtom2} . However, the
behaviour of this discrepancy is not uniform across the periodic table:
for light atoms $\delta g$ is almost constant, whereas in heavy atoms
it changes its sign and increases rapidly with $Z$, suggesting two
different underlying mechanisms. In Ref. \citep{HeavyAtom1, HeavyAtom2} the
authors showed how relativistic corrections and the effect of core
polarization account for the behaviour of $\delta g$ in heavy atoms
such as cesium. Experimental data is available for Cs, Fr, and Rb and
comparison with our results for Cs and Fr yields good agreement
(in  Rb there is a 
cancellation of two contributions which increases the relative error).

\section{Method}
To calculate transition amplitudes and g-factors we utilize the
relativistic Hartree-Fock (Dirac-Fock) approximation in a $V^{N-1}$
potential. Core polarization and core-valence correlations are
included by means of the time-dependent Hartree-Fock (TDHF) and
correlation potential methods~\cite{CPM}.  
The TDHF method is equivalent to well-known random-phase approximation (RPA).

The Hartree-Fock Hamiltonian has the form 
\begin{equation}
	\hat H_0 =  c \alpha \cdot {\bf \hat p} + (\beta-1)mc^2
     - \frac{Ze^2}{r} + \hat V,
\label{HFG}
\end{equation}
where $\hat V$ is the self-consistent potential created by electrons 
from the core. In addition to the Coulomb interaction, we include the
effect of magnetic interactions and retardation via the Breit
interaction, the details of which are given in~\citep{Breit}. 

In the TDHF calculations every single electron wave function of the
atom is presented in the form
\begin{eqnarray}
  \tilde \psi_n &=& \psi_n +X_a e^{-i\omega t} + Y_n e^{i\omega t}
 \label{psi}
\end{eqnarray}
where index $n$ enumerates single-electron states, $\psi_n$ is unperturbed
wave function for the state $n$ which is an eigenstate of the Hartree-Fock
Hamiltonian~(\ref{HFG}), and $X_n$ and $Y_n$ are corrections due to the
magnetic field of an external photon with frequency $\omega$.

These corrections, applicable to all atomic states, are found by
self-consistent iteration of the TDHF equations: 
\begin{eqnarray}
  (\hat H_0 - \epsilon_n - \omega)X_n &=& -(\hat H_{M1} + \delta \hat V_{M1})\psi_n,\nonumber \\
  (\hat H_0 - \epsilon_n + \omega)Y_n &=& -(\hat H_{M1}^{\dagger} + 
  \delta \hat V_{M1}^{\dagger})\psi_n. \label{eq:M1} 
\end{eqnarray}
Here $\delta \hat V$ is the correction to the self-consistent Hartree-Fock
potential $\hat V$ due to the external dipole magnetic (M1) field. The equations in~\eref{eq:M1} are first solved self-consistently for all states in the core. Then corrections to valence states are calculated in the field of frozen core. 

In the relativistic case the matrix elements for the operator
$\hat{H}_{M1} = \vec{\mu}\cdot \vec{B}$ and wavefunctions 
\begin{gather}
\psi(r) =\frac{1}{r} \begin{pmatrix}
    f({r}) \Omega_{\kappa m}   \\
      i \alpha g({r}) \Omega_{-\kappa m}
\end{pmatrix},
\end{gather}
are given by
\begin{align}
\label{eq:MatrixEl}
\langle \psi_a|\hat{H}_{M1}| \psi_b \rangle = (\kappa_a +\kappa_b)&\langle -\kappa_a || C^1 || \kappa_b\rangle \\
&\times \int 3\left(f_a g_b+g_a f_b\right) j_1 (kr)dr, \notag{}
\end{align}
where $\kappa = l$ for $j=l-1/2$, $\kappa =-l-1$ for $j=l+1/2$, $C^1$ is normalized spherical harmonic, wavevector $k = \omega/c$ and $j_l(kr)$ is the spherical Bessel function. 

When core polarization is included the matrix element $\langle a||M1||b\rangle$ becomes $\langle a||M1+\delta V_{M1}||b\rangle$. Calculations of these latter matrix elements with RPA included are given in Table~\ref{t:M1}.


\begin{table}[h]
    \centering
   \caption{Magnetic dipole transition amplitudes }
\begin{ruledtabular}
      \begin{tabular}{c|c|ccc}
      &  &   \multicolumn{3}{c}{M1  $\left( |\mu_B| \times10^{-5}\right)$ }  \\
      Element &  Transition     &HF& RPA&Total      \\ \hline
            Rb           	&5s-6s 	&-1.473 	&2.689 	&1.216  	\\ 	
                         	&5s-4d  	&0.2447 	&0.744 	&1.019		 \\ \hline
             Cs           	&6s-7s 	&-1.652	&15.79 	&14.13 	\\ 	
                         	&6s-5d  	&0.5662 	&11.42 	&11.98		 \\ \hline
             Ba$^+$      	&6s-7s 	&-4.050 	&17.58 	&13.53 	\\ 	
                         	&6s-5d  	&2.006 	&20.05 	&22.06		\\ \hline
             Fr           	&7s-8s 	&-2.491	&179.0	 	&176.5 	\\ 	
                         	&7s-6d  	&0.7374 	&126.2 	&126.9		 \\ \hline
             Ra$^+$         	&7s-8s 	&-5.744 	&190.8 	&185.1 	\\ 	
                         	&7s-6d  	&2.401 	&207.9 	&210.3		 \\ \hline
             Yb$^+$         	&6s-7s 	&-5.536	&48.71 	&43.17 	\\ 	
                         	&6s-5d  	&2.166 	&50.25 	&52.42		\\ \hline
             Ac$^{2+}$   	&7s-8s 	&-8.911 	&-2382 	&-2390 	\\ 	
                         	&7s-6d  	&3.510 	&210.1 	&213.6		 \\ \hline
             Th$^{3+}$    	&7s-8s 	&-13.23 	&-2536 	&-2549 	\\ 	
                         	&7s-6d  	&4.432 	&207.8 	&212.2		\\ 
          
    \end{tabular}
\end{ruledtabular}%
\label{t:M1}
\end{table}

\begin{table}[h]
    \centering
   \caption{Electric quadrupole transition amplitudes in units of $a_{\mathrm{0}}^{2}$. }
\begin{ruledtabular}
      \begin{tabular}{c|c|c|c|c}
          Element &  Transition   &This work  & Other calcs. & Experiment   \\ \hline
            Rb                  	&5s-4d  	& 33.42	&	-&- \\ \hline
             Cs                  	&6s-5d  	& 33.60	&-	&35$\pm$3.5 \citep{CsE2Expt}\\ \hline
             Ba$^+$      	&6s-5d  	&12.69	 	& 12.734\citep{SahooM1E2}&12.74$\pm$ 0.37\citep{BaE2Expt}\\ &&&12.63 \citep{SafronovaE2}\\&&&12.625\citep{Gopakumar}\\&&& 12.74\citep{SahooE2}	&	\\ \hline
             Fr                  	&7s-6d  	&  33.59	&-	&-\\ \hline
             Ra$^+$        	&7s-6d  	& 14.77	&	14.59 \citep{Safronova2007}	&-\\ \hline
             Yb$^+$       	&6s-5d  	&12.19 	&	-&-\\ \hline
             Ac$^{2+}$   	&7s-6d  	& 	9.58	&9.52 \citep{Safronova2007}	&-\\ \hline
             Th$^{3+}$    	&7s-6d  	& 7.10		&-	&-\\ 
          
    \end{tabular}
\end{ruledtabular}%
\label{t:E2}
\end{table}

\begin{table}[t]
    \centering
    \caption{g-factor anomaly $\delta g$ $\left(\times 10^{-5}\right)$ for $s_{1/2}$, $d_{3/2}$, and $d_{5/2}$ states. The g$_{j}$-factor may be recovered using $g_{1/2}~=~\delta g_{1/2}+g_{\mathrm{free}}$, $g_{3/2}~=~\delta g_{3/2}+(6-g_{\mathrm{free}})/5$, and $g_{5/2}~=~\delta g_{5/2}+(g_{\mathrm{free}}+4)/5$ respectively for the three states considered.
Here  $g_{\mathrm{free}} = 2.002319304$ is the measured free electron g-factor \citep{NISTCodata}.}
\begin{ruledtabular}
\begin{tabular}{c|c|c|c|c|c}
             Element    & State &$\delta g_{\mathrm{HF}}$  & $\delta g_{\mathrm{RPA}}$  & $\delta g_{\mathrm{total}}$ & $\delta g_{\mathrm{expt}}$    \\ \hline
Rb 	&	$5s_{1/2}$	&	-2.6	&	4.9	& 	2.3	& 	 1.18$\pm$0.2\citep{CsRbGfactor} \\ 
 	& 	$4d_{3/2}$	&	 -0.4 	&	 -0.9 	&	 -1.3 	&	 - \\
 	& 	$4d_{5/2}$	&	 -0.6 	&	 0.0 	&	 -0.6 	&	 - \\ \hline
 Cs       & 	$6s_{1/2}$	& 	-2.9	&	 28.4	& 	25.5	& 	 22.1$\pm$0.2  \citep{CsRbGfactor}  \\ 
	& 	$5d_{3/2}$	&	 -0.7 	&	 -4.8 	&	 -5.5 	&	 - \\
	& 	$5d_{5/2}$	&	 -1.3 	&	 1.2 	&	 -0.1 	&	 - \\ \hline
Ba$^+$  &	 $6s_{1/2}$	&	 -6.4	&	31.6	&	 25.2	&	17.13$\pm$0.11\citep{BaGfactor6s}\\ 
	&			&		&		&		&	17.29$\pm$0.1 \citep{BaGFactor5D3}\\
	&	 $5d_{3/2}$	&	 -4.7 	&	 -16.7 	&	 -21.4 	&-20.8$\pm$0.03 \citep{BaGFactor5D3}	  \\
	&	 $5d_{5/2}$	&	 -7.3 	&	 1 	&	 -6.3 	&	 -9.29$\pm$0.7~\citep{BaGfactor5D}\\ \hline
Fr 	&	$7s_{1/2}$	& 	-4.3	&	335.7	&	 331.4	& 265.1$\pm$9 \citep{FrExpG}\\ 
	&	 $6d_{3/2}$	&	-0.7 	&	 -26.7 	&	 -27.4 	&	 - \\
	&	 $6d_{5/2}$	&	-1.2 	&	 12.2  	&	 11.0 	&	 - \\ \hline
Ra$^+$ &	 $7s_{1/2}$	& 	-8.8	&	356.2	&	 347.4	&- \\ 
	&	 $6d_{3/2}$	&	 -4.7 	&	 -73.4 	&	 -78.1 	&	 - \\
	&	 $6d_{5/2}$	&	 -6.6 	&	 29.6  	&	 23.0 	&	 - \\ \hline
Yb$^+$&	$6s_{1/2}$	& 	-8.6	&	88.4	&	 79.8	&- \\ 
	&	 $5d_{3/2}$	&	 -5.7 	&	 -23.6 	&	 -29.3 	&	 - \\
	&	 $5d_{5/2}$	&	 -8.2 	&	 12.5  	&	 4.3 	&	 - \\ \hline
Ac$^{2+}$ &	$7s_{1/2}$	& 	-13.3	&	342	&	 328.7	&- \\ 
	&	 $6d_{3/2}$	&	 -9.0 	&	 -78.4 	&	 -87.4 	&	 - \\
	&	 $6d_{5/2}$	&	 -11.7 	&	 28.0 	&	 16.3 	&	 - \\ \hline
Th$^{3+}$ &	$7s_{1/2}$	& 	-18	&	326.5	&	 308.5	&- \\ 
	&	 $6d_{3/2}$	&	 -13.5 	&	 -78.4 	&	 -91.9 	&	 - \\
	&	 $6d_{5/2}$	&	 -16.7 	&	   24.4	&	 7.7 	&	 - \\ 

    \end{tabular}
\end{ruledtabular}%
\label{t:gFactor}
\end{table}

\section{Results}
Tables \ref{t:M1}, \ref{t:E2} and \ref{t:gFactor} contain a summary of our results for all elements considered. In all cases the transition energy is taken as the experimental value \citep{NIST,BWParis}, except for $\mathrm{Ac^{2+}}$ 7s$\to$8s where the transition energy is taken from calculation~\citep{Roberts}. Table \ref{t:M1} clearly illustrates the importance of core polarization for M1 amplitudes. Indeed, in almost all systems the RPA value is several orders of magnitude larger than that given by Hartree-Fock calculations alone. 

The most well studied system in Table~\ref{t:M1} is cesium. Experimental values of the reduced  6s-7s M1 amplitude are the following:
\begin{align}
M &= (9.04\pm 0.588)\times 10^{-5}|\mu_B| \textrm{\cite{CsM1a}}  \notag{} \\
M &= (10.1\pm 0.441)\times 10^{-5}|\mu_B| \textrm{\cite{CsM1b}}  \notag{} \\
M &= (10.3\pm 0.196)\times 10^{-5}|\mu_B| \textrm{\cite{CsM1c}} 
\end{align}
compared with the value presented here of $M = 14.13\times 10^{-5}|\mu_B|$.

There is another contribution to the M1 amplitude - the nuclear-spin-dependent (NSD) amplitude  induced by the hyperfine interaction. It  can be separated experimentally due to its dependence on the hyperfine component  of the M1 transition 
In $s-s$ transitions the hyperfine induced amplitude is an order of magnitude smaller than the nuclear-spin-independent (NSI) amplitude (see e.g. \cite{DzubaPRA2000}).
In the $s-d$ transitions the NSD amplitude is even  smaller since the non-diagonal $s-d$ hyperfine interaction matrix elements are smaller.

In addition to M1 amplitudes, we present E2 $s-d$ transition amplitudes given in Table~\ref{t:E2}. In these calculations we have included the effect of correlations using the all-order correlation potential $\hat{\Sigma}$ method (see e.g. \citep{CPM}). Applying this method to M1 amplitudes makes little difference as $\hat{\Sigma}$ changes only the radial wavefunction, which M1 transitions are not sensitive to. 

The forbidden M1 amplitudes are very small and sensitive to different
corrections. Matrix elements of the M1 operator are very sensitive to
the frequency of the laser field $\omega$. All g-factors are calculated at
$\omega=0$. This is why they are similar. In contrast, each M1
amplitude is calculated at the frequency of this transition and
frequencies grow rapidly with the degree of ionization. This is why
the amplitudes are different. For example,
the frequency of the 7s-8s transition in Ac$^{2+}$ is about 2 times
larger than in Ra$^+$. If we calculate the M1 amplitude for Ac$^2+$ at
the same frequency as in Ra$^+$, we get an answer very close to that
of Ra$^+$. 

Our method to estimate the higher order corrections to M1
amplitudes that are omitted is based on the calculations of the
g-factor anomalies which have similar mechanisms. Results of our
calculations for the g-factor anomalies due to the relativistic and
many-body corrections are  presented in Table~\ref{t:gFactor}. 
 Comparison with the experimental data for the g-factor anomalies (and
 6s-7s M1 amplitude in Cs)  indicates that the theoretical error in
 our calculation is from 10 to  40 \% .  

Previous calculations for $s-d$ M1 transitions exist for some systems:
$80\times 10^{-5}|\mu_B|$~\citep{SahooM1E2} for Ba$^+$, and $140\times
10^{-5}|\mu_B|$ and $130\times 10^{-5}|\mu_B|$ for Ra$^+$ and
Ac$^{2+}$ respectively~\citep{Safronova2007}. The values for Ra$^+$
and Ac$^{2+}$, while differing from our calculations given in
Table~\ref{t:M1}, are nevertheless consistent with our values given
the error estimates discussed previously. In the case of Ba$^+$ the
difference is too large and should be treated as disagreement between
present calculations and those of Ref.~\citep{SahooM1E2}.
Additionally, the calculations of the E2 transition amplitudes
for these ions have also been performed previously. In
the case of Ba$^+$ experimental data for E2 exists and a comparison with our
result indicates an accuracy of better than 1\%. Comparison may also
be made with previous calculations, the results of which are given in
Table~\ref{t:E2}, which are consistent with this level of accuracy.  



\section*{Acknowledgements}
V. F. is grateful to the Humboldt Foundation for their support, and to the Frankfurt Institute for Advanced Study for their hospitality. This work was supported in part by the Australian Research Council.

The authors are also grateful to M. R. Hoffman for bringing the important experimental results in \citep{BaGFactor5D3} to our attention.

\end{document}